# Embedded Ferroelectric Nanoclusters can drive Polarization Reversal in a Non-Ferroelectric Polar Film via the Proximity Effect

Anna N. Morozovska[1], Eugene A. Eliseev[2], Sergei V. Kalinin[3], Long-Qing Chen[4*],
Dean R. Evans[5†], and Venkatraman Gopalan[4‡]

[1] Institute of Physics of the National Academy of Sciences of Ukraine,
46, Nauki Avenue, 03028 Kyiv, Ukraine

[2]Frantsevich Institute for Problems in Materials Science, National Academy of Sciences of Ukraine,
3, str. Omeliana Pritsaka, 03142 Kyiv, Ukraine

[3] Department of Materials Science and Engineering, University of Tennessee, Knoxville, TN, 37996, USA

[4] Department of Materials Science and Engineering,
Pennsylvania State University, University Park, PA 16802, USA

[5]Zone 5 Technologies, Special Projects, San Luis Obispo CA  93401, USA

## Abstract

Heterogeneous nucleation from defects dominates the electric field required for polarization switching of ferroelectrics. Here, we consider the switching of a nominally non-switchable polar thin film of AlN due to the proximity effect arising from embedded ferroelectric nanoclusters of $Al_{1-x}Sc_xN$. Using a Landau-Ginzburg-Devonshire thermodynamic approach and finite element modeling, we study the influence of nanocluster shape on polarization switching and domain nucleation emerging in AlN. The ferroelectric nanocluster boundary is modeled as a thin layer transitioning from $Al_{1-x}Sc_xN$ to AlN. We analyze the conditions under which polarization switching in the AlN film occurs at coercive fields significantly lower than its dielectric breakdown field. In the presence of spike-like $Al_{1-x}Sc_xN$ nanoclusters, the proximity effect enables switching of the spontaneous polarization in AlN and significantly reduces the corresponding coercive field. The internal field, which is depolarizing inside the AlN (due to its larger spontaneous polarization) and polarizing within the ferroelectric $Al_{1-x}Sc_xN$ nanoclusters (due to its smaller spontaneous polarization), lowers the potential barrier in the clusters and nucleates nanodomains at the $Al_{1-x}Sc_xN$–AlN interface, forming localized regions of reversed polarization. Proximity effect can thus provide a pathway towards "thawing" previously "frozen" ferroelectrics through engineered nucleation for memory, actuation and optical technologies.

[*] corresponding author, e-mail: lqc3@psu.edu

[†] corresponding author, e-mail: deanevans@zone5tech.com

[‡] corresponding author, e-mail: vgopalan@psu.edu, vxg8@psu.edu

## 1. INTRODUCTION

Nanoscale ferroelectrics with wurtzite and fluorite structure are of great scientific interest due to their unique ferroelectric, dielectric, and piezoelectric properties [1, 2, 3]. Recently discovered ferroelectric nitrides and oxides with the wurtzite structure, such as nanoscale *lead-free* $Al_{1-x}B_xN$, $Al_{1-x}Sc_xN$, $Al_{1-x}Hf_xN$, $Zn_{1-x}Mg_xO$, and fluoride $Hf_xZr_{1-x}O_2$, are considered to be among the most promising candidates for the next generation of Si-compatible and easily integrable electronic memory elements, such as ferroelectric random access memory (FeRAM), steep-slope field-effect transistors, various logic devices, and piezoelectric actuators [1-3]. These solid-state materials are also chemically and thermally stable, environmentally friendly, and characterized by relatively low cost and simple synthesis methods.

The electric polarization switching in nitride ferroelectrics is hysteretic with a very large spontaneous polarization (up to 1.2 – 2.0 $\mu C/cm^2$), but at the same time it also has a very large coercive field, which reaches from 5 to 15 MV/cm for nitrides like $Al_{1-x}Sc_xN$ [4, 5]. These coercive field values are close to the dielectric breakdown fields or exceed them, sometimes by several fold. According to density functional theory (DFT) calculations, the free energy of single-domain uniaxial nitride ferroelectrics has extremely deep and widely separated potential wells, which correspond to opposite directions of spontaneous polarization, separated by a high potential barrier [6]. The DFT results correspond to unrealistically large coercive fields, because the domain nucleation barrier depends on the domain wall energy, which is related to the depth of the potential wells. However, in ferroelectric wurtzites, the electrostatic and elastic energies can play a dominant role in determining the domain nucleation barrier [7].

It is known that atomic-scale chemical stresses induced by bulk doping can stabilize a ferroelectric phase in fluorites (as e.g., in Zr doping of $HfO_2$) [8]. Furthermore, the chemical strain associated with doping can significantly reduce the height of the free energy potential barriers and thus lower the coercive field to practically acceptable values [9, 10, 11]. For example, the barrier is lowered both when doping the polar but non-switchable piezoelectric zinc oxide (ZnO) with magnesium [12] and when doping the polar but non-switchable aluminum nitride (AlN) with scandium or boron [13, 14]. However, doping, as a rule, has a detrimental effect on other functional properties of wurtzite ferroelectrics, such as dielectric losses, electrochemical activity [15], and significant optical scattering losses that impede optoelectronic and other applications [16].

Emerging ideas aim to exploit both size effects and “proximity to the ferroelectric” in otherwise non-switchable polar materials to induce ferroelectric switching of AlN and ZnO piezoelectrics without electrical breakdown [17]. Polarization reversal has been experimentally observed in AlN and ZnO piezoelectric layers that were in direct electric contact with “chemically related” ferroelectric layers, such as $Al_{1-x}B_xN$, $Al_{1-x}Sc_xN$, and $Zn_{1-x}Mg_xO$, in multilayer films. The layered structures, whose thicknesses varied from tens to hundreds of nm, included two-layer (asymmetric, e.g., $Al_{1-x}Sc_xN$/AlN, $Al_{1-x}B_xN$/AlN, ZnO/$Al_{1-x}B_xN$) and three-layer (symmetric, e.g., $Al_{1-x}B_xN$/AlN/$Al_{1-x}B_xN$, AlN/$Al_{1-x}B_xN$/AlN, $Zn_{1-x}Mg_xO$/ZnO/$Zn_{1-x}Mg_xO$) configurations [17]. The first attempts of the “proximity to the ferroelectric" thermodynamic description in wurtzite multilayer structures are presented in Refs. [18, 19].

Compositionally graded ferroelectric materials can be regarded as the natural limit of multilayer structures as the layer thickness decrease. Prof. Alpay group [20, 21, 22, 23] developed a generalized Landau-Ginzburg-Devonshire (LGD) approach for analyzing compositionally graded ferroelectric nanomaterials, revealing the leading role of compositional strain and stress gradients in the formation of domain structures and polarization reversal under an applied electric field. As was recently discovered [24], the compositionally graded planar structures AlN-$Al_{1-x}Sc_xN$ and ZnO-$Zn_{1-x}Mg_xO$ allow simultaneous switching of spontaneous polarization throughout the system by a coercive field significantly lower than the electric breakdown field of unswitchable polar materials. This simultaneous switching is governed by a depolarization electric field resulting from the gradient of the chemical composition parameter "x" (Sc or Mg), which reduces the switching barrier in the otherwise unswitchable regions of the compositionally graded structures [24].

However, to the best of our knowledge, polar and electrophysical properties of compositionally graded wurtzite nanoclusters and nanoparticles have not been studied. Methods for controlling these properties by changing the chemical composition gradient, shape, and size effects, and/or proximity effects have not been developed. There is scientific and practical interest for creating silicon-compatible wurtzite nanomaterials with high spontaneous polarization and significantly reduced coercive fields that can be controlled by chemical and physical factors, which can be very promising for advanced nanoelectronics, optoelectronics, and related emerging technologies.

To fill the gap in knowledge, we theoretically analyze the influence of nanocluster shape on the polarization switching and domain nucleation emerging in otherwise non-switchable polar films due to the proximity of ferroelectric nanoclusters. To describe this system, the boundary of the ferroelectric nanocluster within the non-ferroelectric polar film is modeled as a compositionally graded layer. Using a LGD thermodynamic approach and finite element modeling (FEM), we analyze the conditions that allow switching the electric polarization of the AlN film at the coercive field significantly lower than the electric breakdown field due to the proximity of ferroelectric $Al_{1-x}Sc_xN$ nanoclusters. We also explore the underlying physical mechanisms of the proximity effects in the non-ferroelectric films with embedded ferroelectric nanoclusters.

## 2. PROBLEM FORMULATION

Let us consider switchable ferroelectric $Al_{1-x}Sc_xN$ nanoclusters embedded in the otherwise non-switchable AlN film covered by conducting electrodes (shown in **Fig. 1(a)**). The XZ cross-section of the clusters has the width $2R$ and the height $d$. The clusters are modeled as wires extending along the Y direction, which allows us to solve a quasi-2D problem. The boundary between the $Al_{1-x}Sc_xN$ clusters and the AlN material is a compositionally graded layer, whose thickness is determined by the diffusion length $\Delta$. The thickness of the AlN film is $h$. The period $L$ between the clusters coincides with the lateral size of the computational cell. For comparison, we consider the inverted structure consisting of AlN nanoclusters in an $Al_{1-x}Sc_xN$ film covered by electrodes (shown in **Fig. 1(b)**).

An electric voltage $U(t)$ is applied to the top electrode (its electric potential is $\varphi = U(t)$). The voltage amplitude increases linearly in time from zero to $U_{max}$, such that $U(t) = U_{max}\frac{t}{t_{max}}\sin(\omega t)$, where $\omega$ is the pulse frequency and $t_{max}$ is the computation time. Such form of the voltage sweep corresponds to existing experiments [12, 15] and allows us to study different stages of polarization reversal [18, 19, 24] in the clusters and in the film. The bottom electrode is regarded as electrically grounded (its electric potential is $\varphi = 0$).

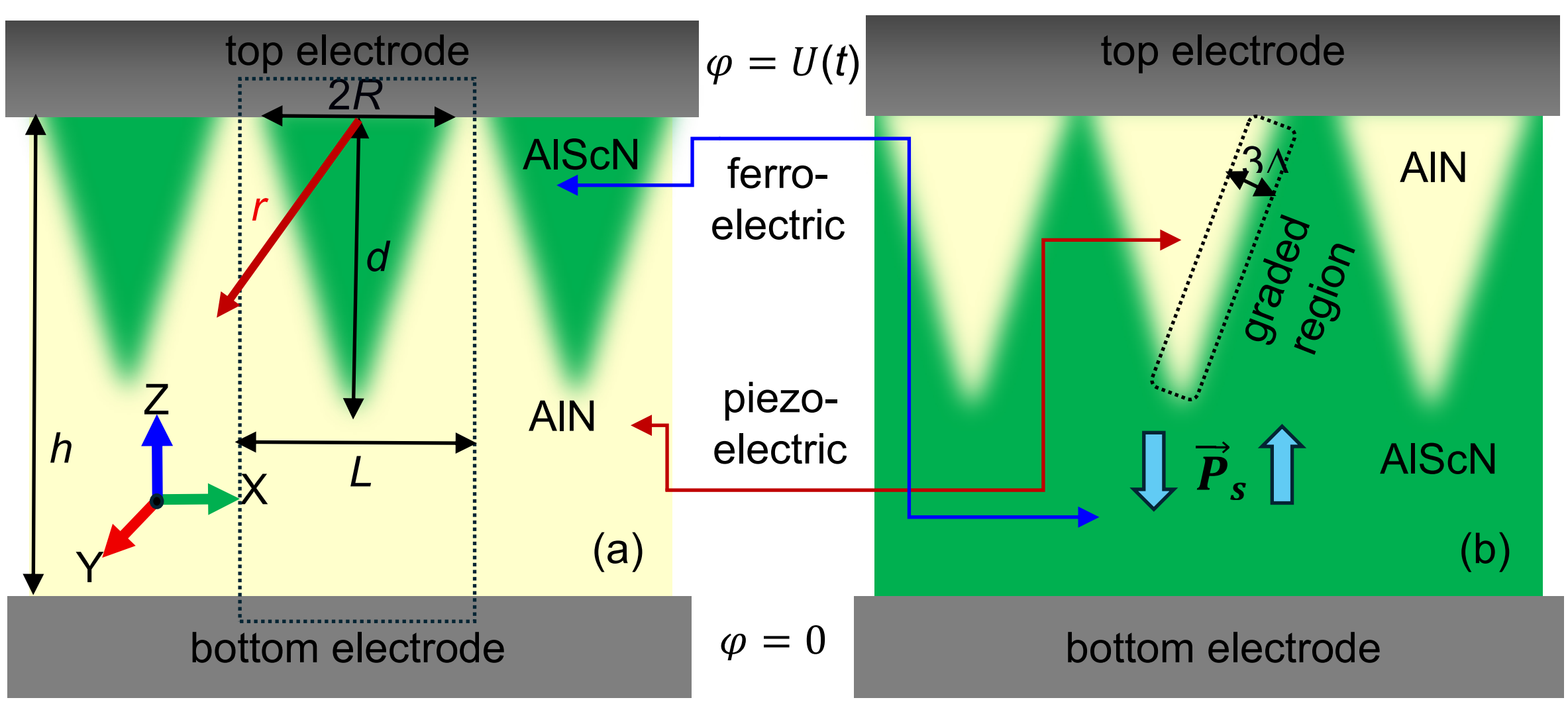

**FIGURE 1**. **(a)** $Al_{1-x}Sc_xN$ nanoclusters in an AlN film, and **(b)** AlN nanoclusters in an $Al_{1-x}Sc_xN$ film, placed between parallel-plate electrodes. The boundary between $Al_{1-x}Sc_xN$ and AlN is a compositionally graded layer with an effective thickness of approximately $3\Delta$, where $\Delta$ is the diffusion length. The cluster width is $2R$ and its height is $d$; $\vec{r}$ is the radius-vector counted from the cluster center. The clusters are regarded as wires extending along the Y direction, which is perpendicular to the figure plane. The film thickness is $h$. The polar c-axis of the wurtzite structure is normal to the electrode surfaces and coincides with Z-axis. A dotted rectangle in part **(a)** shows the lateral size $L$ of the computational cell. Two directions of the out-of-plane spontaneous polarization $\vec{\boldsymbol{P}}_s$ are shown by thick arrows in part **(b)**.

Our goal is to determine the conditions (e.g., the optimal nanocluster shape, dimensions $R$, and $d$, diffusion length $\Delta$, and the cluster period $L$) that allows switching the spontaneous polarization of the non-ferroelectric polar film, via the proximity effect, at a coercive field significantly lower than the dielectric breakdown field. We also investigate the emergence of the proximity effect in the inverted structure for different values of $R$, $d$, $\Delta$, and $L$.

The time-dependent LGD equation for the ferroelectric polarization $P_z$ inside a compositionally graded wurtzite nanocluster can be written as:

$$\Gamma\frac{\partial P_z}{\partial t} + \alpha(\vec{r})P_z + \beta(\vec{r})P_z^3 + \gamma(\vec{r})P_z^5 - g_z\frac{\partial^2 P_z}{\partial z^2} - g_\perp\Delta_\perp P_z = E_z. \quad (1a)$$

Here, $\Gamma$ is the Landau-Khalatnikov relaxation coefficient, $\Delta_{\perp}$ is the transverse part of Laplace operator, and $E_z = -\frac{\partial \varphi}{\partial z}$ is z-component of electric field. The Landau expansion coefficients $\alpha$, $\beta$, and $\gamma$ are coordinate-dependent due to the gradient of the chemical composition "$x$" in the following way:

$$\alpha(\vec{r}) = \alpha_1 + (\alpha_2 - \alpha_1) f(\vec{r}) - 2Q_{1133}(\sigma_{22} + \sigma_{11}) - 2Q_{3333}\sigma_{33}, \tag{1b}$$

$$\beta(\vec{r}) = \beta_1 + (\beta_2 - \beta_1) f(\vec{r}), \tag{1c}$$

$$\gamma(\vec{r}) = \gamma_1 + (\gamma_2 - \gamma_1) f(\vec{r}). \tag{1d}$$

The coefficients $\alpha_1$, $\beta_1$, and $\gamma_1$ correspond to the ferroelectric $Al_{0.73}Sc_{0.27}N$; the coefficients $\alpha_2$, $\beta_2$, and $\gamma_2$ correspond to the AlN structure shown in **Fig. 1(a)** (or vice versa for the structure shown in **Fig. 1(b)**). The Fermi-type function $f(\vec{r})$ is introduced as

$$f(\vec{r}) = \begin{cases} 0, \text{ inside the cluster,} \\ \text{changes from 0 to 1 in the graded layer,} \\ 1, \text{ outside the cluster.} \end{cases} \tag{1e}$$

Here $\vec{r}$ is the radius-vector counted from the cluster center, the radius-vector $\vec{r}_S$ describe its surface, and the diffusion length $\Delta$ determines the thickness of the graded layer, where $|\vec{r} - \vec{r}_S| \leq 3\Delta$.

The coefficient $\alpha(\vec{r})$ is also renormalized by elastic stresses $\sigma_{ij}$ via the electrostriction coefficients $Q_{ijkl}(\vec{r})$. Hereafter, we assume that both elastic compliances $s_{ijkl}(\vec{r})$ and electrostriction coefficients $Q_{ijkl}(\vec{r})$ are coordinate-dependent in the same functional way as the Landau expansion coefficients, namely, $s_{ijkl}(\vec{r}) = s_{ijkl}^{(1)} + \left(s_{ijkl}^{(2)} - s_{ijkl}^{(1)}\right) f(\vec{r})$ and $Q_{ijkl}(\vec{r}) = Q_{ijkl}^{(1)} + \left(Q_{ijkl}^{(2)} - Q_{ijkl}^{(1)}\right) f(\vec{r})$, where the tensor components $s_{ijkl}^{(1)}$ and $Q_{ijkl}^{(1)}$ correspond to the ferroelectric $Al_{0.73}Sc_{0.27}N$; the coefficients $s_{ijkl}^{(2)}$ and $Q_{ijkl}^{(2)}$ correspond to the AlN structure shown in **Fig. 1(a)** (or vice versa for the structure shown in **Fig. 1(b)**).

Elastic stresses satisfy the equation of mechanical equilibrium in the computation region, $\frac{\partial \sigma_{ij}}{\partial x_j} = 0$. Elastic equations of state follow from the variation of the free energy with respect to elastic stress, and are given by:

$$s_{ijkl}\sigma_{ij} + Q_{ijkl}P_k P_l = u_{ij}. \tag{2}$$

Elastic boundary conditions correspond to the absence of normal stress at the top electrode ($z = h$) and zero elastic displacement at the bottom electrode ($z = 0$) assuming that the film is clamped to a rigid substrate. Hereafter, we neglect the influence of the flexoelectric coupling for the sake of simplicity.

For this work, we assume that the polarization and its derivatives are continuous functions inside the nanostructured film. The boundary conditions for polarization at the top ($z = h$) and bottom ($z = 0$) surfaces of the film are of the third type [25]:

$$\left. \frac{\partial P_z}{\partial z} - \frac{P_z}{\lambda_1} = 0 \right|_{z=0}, \qquad \left. \frac{\partial P_z}{\partial z} + \frac{P_z}{\lambda_2} = 0 \right|_{z=h}, \tag{3}$$

where $\lambda_1$ and $\lambda_2$ are the so-called extrapolation lengths [26], whose values depend on the surface-electrode pair, the nature of the short-range interactions at the surface/interface, and/or the preparation conditions; these quantities serve as fitting parameters. The limiting case $\lambda_i \to 0$ corresponds to zero polarization at the surface and favors domain formation under incomplete screening conditions. The limiting case $\lambda_i \rightarrow \infty$, used in this work, corresponds to the so-called natural boundary conditions, which provide the strongest stabilization of the initial single-domain state of the thin film covered with ideally conducting electrodes.

The electric potential $\varphi$ obeys the Poisson equation, $\varepsilon_0 \frac{\partial}{\partial x_i}\left[\varepsilon_{ij}(\vec{x}) \frac{\partial}{\partial x_j} \varphi(\vec{x})\right] = \frac{\partial P_k(\vec{x})}{\partial x_k}$. Since the wurtzite structures are considered as uniaxial ferroelectrics with the polar axis Z, the in-plane components of polarization take the form:

$$P_1(\vec{r}) = -\varepsilon_0[\varepsilon_{11}(\vec{r}) - 1]\frac{\partial \varphi}{\partial x}, \qquad P_2(\vec{r}) = -\varepsilon_0[\varepsilon_{11}(\vec{r}) - 1]\frac{\partial \varphi}{\partial y}, \tag{3a}$$

where the function $\varepsilon_{11}(\vec{r}) = \varepsilon_{11}^{(1)} + \left(\varepsilon_{11}^{(2)} - \varepsilon_{11}^{(1)}\right)f(\vec{r})$; $\varepsilon_{11}^{(1)}$ and $\varepsilon_{11}^{(2)}$ are in-plane dielectric permittivities, $\varepsilon_0$ is a universal dielectric constant. The out-of-plane component of polarization has the form:

$$P_3(\vec{r}) = P_z(\vec{r}) - \varepsilon_0[\varepsilon_b(\vec{r}) - 1]\frac{\partial \varphi}{\partial z}, \tag{3b}$$

where $P_z(\vec{r})$ is the ferroelectric/piezoelectric part of polarization related with the soft mode; the function $\varepsilon_b(\vec{r}) = \varepsilon_b^{(1)} + \left(\varepsilon_b^{(2)} - \varepsilon_b^{(1)}\right)f(\vec{r})$; $\varepsilon_b^{(1)}$ and $\varepsilon_b^{(2)}$ correspond to the background permittivity [27] of ferroelectric and the non-ferroelectric uniaxial polar materials, respectively. Note that the division on the soft-mode and background contributions is relevant for the polar direction Z only.

From Eqs.(3a) and (3b), the Poisson equation acquires the form:

$$\varepsilon_0\left[\varepsilon_{11}(\vec{r})\Delta_\perp \varphi + \frac{\partial}{\partial z}\left(\varepsilon_b(\vec{r})\frac{\partial \varphi}{\partial z}\right)\right] = \frac{\partial P_z}{\partial z}. \tag{4a}$$

The electric boundary conditions are the fixed potential at the electrodes,

$$\varphi|_{z=0} = 0, \qquad \varphi|_{z=h} = U(t). \tag{4b}$$

LGD parameters of $Al_{0.73}Sc_{0.27}N$ and AlN, used in our calculations, are listed in **Table S1** in Supplementary Materials [28]. They are determined from Refs. [4, 5, 29, 30]. Electrostriction coefficients $Q_{ijkl}$ and elastic stiffness tensors $c_{ijkl}$, listed in **Tables S2** [28], are taken from Ref. [31]. The procedure for determining the LGD parameters solving the constitutive equations FEM is described in detail in Refs. [18, 19, 24].

We perform FEM calculations in COMSOL@MultiPhysics software, using the electrostatics, solid mechanics, and general math (PDE toolbox) modules, with varying rectangular mesh discretization densities and polarization relaxation conditions. The maximal size of the quasi 2D-computational region is L×h×D nm$^3$, where the distance D in the Y-direction can be chosen small enough due to the independence of all physical properties on the coordinate "y".

The diffusion length Δ varies from 0.5 to 3 nm, and the period $L$ is taken to be several times larger than the cluster size. Examples of relatively sharp and diffuse stripes and semi-ellipsoidal clusters used in our calculations are shown in **Fig. S1** in **Supplementary Materials**.

The semi-ellipsoidal shape of the cluster XZ cross-section is chosen because its depolarization field factor $n$ is given by the simple analytical expression, $n = \frac{R}{R+d}$, where $R$ and $d$ are the semiaxes of the ellipse along the X and Z directions (see e.g. Ref. [32]). The factor $n$ varies from 0 (for vertical stripes) to 1 (for horizontal layers), and equals 0.5 for a semi-sphere. The depolarization factor is determined by the cluster geometry and can be expressed in terms of the aspect ratio $\eta = \frac{R}{d}$ as $n = \frac{\eta}{\eta+1}$. The depolarization factor determines the internal depolarizing field $E_d$ inside the uniformly polarized cluster, which is proportional to $\frac{n}{\varepsilon_0}\Delta P_z$, where $\Delta P_z$ denotes the difference of the spontaneous polarizations outside and inside the cluster. The field $E_d$ is inhomogeneous outside the cluster and, as a rule, is sign-alternating. When $\Delta P_z < 0$, the internal field $E_d$ inside the cluster is negative and "depolarizing", whereas when $\Delta P_z > 0$, the field $E_d$ inside the cluster is positive and "polarizing". The field outside the cluster is often called a "stray field" due to its inhomogeneity, and it weakens with distance from the cluster.

Periodic boundary conditions are imposed in the Y-direction and at lateral boundaries of the computational cell. The average size of the mesh element is equal to 0.5 nm. The physical properties dependence on the mesh size is verified by increasing the size to 1.0 nm. We found that this results in minor changes in the electric polarization, electric field, and elastic stresses and strains, such that the spatial distribution of each of these quantities becomes less smooth. When using these larger cell sizes, all significant details remain visible, and, more importantly, the system free energy saturates as the mesh size decreases below 1 nm.

## 3. RESULTS AND DISCUSSION

To find the conditions that allow switching of the spontaneous polarization in the AlN film at coercive fields significantly lower than the dielectric breakdown field, we vary the shape of $Al_{0.73}Sc_{0.27}N$ clusters by changing the aspect ratio $\eta$ while maintaining a fixed area of the semi-ellipsoidal cluster cross-section, $S = \pi R d/2$ (see schematics in **Fig. 2(a)**). To study the role of the clusters as polar defects, we consider the case when their relative volume fraction $v_c = S/(hL)$ is small, such that $v_c \leq 0.1$. To study the role of diffusion lengths, we compare the results calculated for small $\Delta = 0.5$ nm (in the main text) and relatively large $\Delta = 3$ nm (in the Supplement [28]).

For comparison with the influence of semi-ellipsoidal clusters, we consider the polarization switching in an $Al_{0.73}Sc_{0.27}N$/AlN bilayer structure separated by a compositionally graded layer, as well as in a vertical $Al_{0.73}Sc_{0.27}N$–AlN striped structure, which are shown schematically in **Fig. 2(b)** and **2(c)**, respectively. The thickness of the $Al_{0.73}Sc_{0.27}N$ layer $d$ and the width of $Al_{0.73}Sc_{0.27}N$ stripes $w$ are chosen such that their volume fractions, $v_{bl} = \frac{d}{h}$ and $v_{st} = \frac{wh}{L^2}$, are the same as $v_c$. The influence of the diffusion length on the volume fractions is negligibly small for $\Delta \leq 0.5$ nm.

**Figure 2(d)** shows the polarization-voltage hysteresis loops $\overline{P_z}(U)$ calculated for different cross-section shapes of $Al_{0.73}Sc_{0.27}N$ nanoclusters in the AlN film, and **Fig. 2(e)** shows the $\overline{P_z}(U)$ loops calculated for different cross-section shapes of AlN nanoclusters in the $Al_{0.73}Sc_{0.27}N$ film. The dependence of coercive field on the depolarization factor $n$ is shown in **Fig. 3(a)**, where the red curve corresponds to $Al_{0.73}Sc_{0.27}N$ nanoclusters in the AlN film, and the blue curve corresponds to AlN nanoclusters in the $Al_{0.73}Sc_{0.27}N$ film. **Figures 3(b)-(d)** illustrate domain nucleation and vertical and lateral growth states for typical cases ($n \ll 1, n = 0.5$, and $n \sim 1$). **Figure 4(a)** illustrates the polarization distribution near the elongated semi-ellipsoidal $Al_{0.73}Sc_{0.27}N$ nanocluster in the AlN film, and **Fig. 4(b)** corresponds to the inverted structure. The dependence of the internal field z-component $E_d$ on the depolarization factor $n$, calculated near the cluster-electrode interface (points 1 - 3) and near the cluster apex outer boundary (points $2 - 4$) is shown in **Figs. 4(c)** for the $Al_{0.73}Sc_{0.27}N$ nanocluster in the AlN film, and in **Fig. 4(d)** for the inverted structure. **Figure 5** illustrates the features of polarization switching in the structure consisting of spike-like $Al_{0.73}Sc_{0.27}N$ nanoclusters ($R \ll d$, $n \ll 1$) periodically spaced in the AlN film. **Figure 6** illustrates the features of polarization switching in the inverted structure consisting of semi-circular AlN nanoclusters ($R = d$, $n = 0.5$) periodically spaced in the $Al_{0.73}Sc_{0.27}N$ film.

Results, shown in the various panels of **Figs. 2-6**, are calculated in the case of a single-domain initial distribution of polarization in nanostructured films. Specifically, the initial state was an upward-directed spontaneous polarization with randomly small fluctuations in all layers and the boundary condition $\varphi = 0$ imposed at both film surfaces. Note that under ideal surface screening conditions the single-domain state of the spontaneous polarization is the ground state of the nanostructured films. After the spontaneous polarization relaxed to the single-domain state, the voltage was applied using the same time-dependent voltage profile shown in **Figs. 5(a)** and **6(a)**.

Due to the proximity effect, the spontaneous polarization switches simultaneously throughout the film for all considered clusters geometries, while the corresponding coercive fields strongly depend on geometry. From a comparison of the width of $P_3(U)$ loops in **Fig. 2(d)** and from the coercive field dependence on the depolarization factor shown in **Fig. 3**, one can conclude the following: the inclusion of $Al_{0.73}Sc_{0.27}N$ stripes results in the largest reduction of the coercive field in the AlN film (by a factor of ~2.61 compared to pure AlN film). Such strong reduction means that the inclusion of thin $Al_{0.73}Sc_{0.27}N$ stripes can decrease the coercive field of AlN film below its dielectric breakdown field, which can be as high as $10 - 15$ MV/cm [33]. The inclusion of spike-like $Al_{0.73}Sc_{0.27}N$ clusters (with $\frac{d}{R} \geq 6$) reduces the coercive field of the AlN film by a factor of 1.86 compared to pure AlN film. The inclusion of semi-circular $Al_{0.73}Sc_{0.27}N$ clusters leads to a smaller reduction of the coercive field; and the inclusion of flattened $Al_{0.73}Sc_{0.27}N$ clusters (or horizontal layer) results in the weakest reduction of the coercive field (by a factor of ~1.04 compared to AlN film). The remanent polarization, $P_r \approx \pm 122$ μC/cm$^2$, is close to the polarization of bulk AlN. The magnitude of $P_r$ is nearly independent of the cluster shape, because the relative volume fraction $v_c$ of the clusters is small, namely $v_c \approx \pi/40$.

Let us assume that the difference in coercive fields, shown by the red curve in **Fig. 3(a)**, is largely determined by the nucleation of needle-like nanodomains in much less coercive $Al_{0.73}Sc_{0.27}N$ clusters (including the regions very close to their surface), followed by their vertical growth through the film. Meanwhile, the subsequent lateral motion of uncharged domain walls in the AlN matrix is only weakly dependent on the cluster shape. The assumption, which is fully consistent with the FEM results, shown in **Fig. 3(b)-3(c)**, allows us to explain the coercive field behavior in AlN films with $Al_{0.73}Sc_{0.27}N$ clusters in the following way. The coercive field of the AlN films with $Al_{0.73}Sc_{0.27}N$ stripes is the smallest, because the vertical growth of the domain filament occurs entirely within the $Al_{0.73}Sc_{0.27}N$ stripes between the electrodes. The needle-like nanodomains should propagate a certain distance in the polar direction Z in the highly-coercive AlN to reach the bottom electrode. Since the volume fraction of the clusters was chosen to be the same for all cluster shapes, the distance, that must be passed by the domain propagating along Z-direction in AlN to bridge the gap between the electrodes, is largest for the case of the AlScN−AlN bilayer. It is large enough (but a bit smaller) for flattened $Al_{0.73}Sc_{0.27}N$ clusters, smaller for the case of semi-circular $Al_{0.73}Sc_{0.27}N$ clusters, and the smallest for the case of spike-like $Al_{0.73}Sc_{0.27}N$ clusters. The distance is zero for the case of $Al_{0.73}Sc_{0.27}N$ stripes in the AlN film. Thus, the coercive field is largest in the AlScN−AlN bilayer, slightly smaller in the AlN film with flattened $Al_{0.73}Sc_{0.27}N$ clusters, smaller in the AlN film with semi-circular $Al_{0.73}Sc_{0.27}N$ clusters, significantly smaller in the AlN film with spike-like $Al_{0.73}Sc_{0.27}N$ clusters, and smallest in the AlN film with vertical $Al_{0.73}Sc_{0.27}N$ stripes.

The above considerations explain the monotonic increase of the coercive field $E_c$ with an increasing depolarization factor $n$, shown by the red curve in **Fig. 3(a)**, but they do not explain the mechanism of domain nucleation and growth in the nanostructured film. According to the FEM results shown in **Fig. 3(b)-3(c)** and **Fig. 5**, nucleation starts in the spatial regions where the electric field most strongly promotes it. The field is a superposition of the applied field and the internal field $E_d$. The sources of the internal field are the uncompensated bound charges related to the compositional x-gradient. The magnitude of the bound charge is determined by the tilt of the cluster surface with respect to the spontaneous polarization vector (directed along axis Z). The internal and stray fields are absent for the vertical $Al_{0.73}Sc_{0.27}N$ stripes (i.e., for $n = 0$). The side walls of the semi-ellipsoidal $Al_{0.73}Sc_{0.27}N$ clusters are nearly parallel to the polarization vector near the top surface, and thus free of bound charge (similarly to weakly charged domain walls). The bound charge is concentrated near the bottom surface of the clusters, where the polarization vector is normal to the surface (see e.g., **Fig. 4(a)**). The corresponding internal field is quasi-uniform and positive (i.e., polarizing) inside the uniformly polarized semi-ellipsoidal $Al_{0.73}Sc_{0.27}N$ clusters, and its magnitude increases strongly with an increase in the depolarization factor $n$ (see **Figs. 2(f)-2(h)** and curves 1 and 3 in **Fig. 4(c)**). The field is negative and reaches minimum at the bottom surface of the clusters for $n < 0.45$, then it changes sign and increases with an increase in the $n$ value (see curve 2 in **Fig. 4(c)**). The internal stray field is negative and inhomogeneous outside the $Al_{0.73}Sc_{0.27}N$ clusters (see **Figs. 2(f)-2(h)** and curve 4 in **Fig. 4(c)**). The negative depolarization field, which is strongly concentrated in the immediate vicinity outside the sharp apex of the $Al_{0.73}Sc_{0.27}N$ spike-like cluster and much less concentrated outside the flattened $Al_{0.73}Sc_{0.27}N$ cluster (see **Figs. 2(f)-2(h)** and **4(c)** for details), supports the nucleation and vertical growth of needle-like nanodomains in the AlN with a strength that depends on the field concentration. Therefore, the needle-like nanodomains shown in **Figs. 3(b)-3(c)** nucleate along the line connecting the points 2 and 4 in **Fig. 4(a).**

The internal field, caused by the uncompensated bound charges, emerges to minimize the difference in spontaneous polarization of the $Al_{0.73}Sc_{0.27}N$ ($P_{AlScN}$ ≈105 μC/cm$^2$) and AlN ($P_{AlN}$ ≈126 μC/cm$^2$) materials [18, 19, 24]. The negative internal field is "depolarizing" inside the AlN film, because its direction is opposite to the positive polarization of AlN. In contrast, the positive internal field is "polarizing" in the $Al_{0.73}Sc_{0.27}N$ material, because it has the same direction as the initial polarization of $Al_{0.73}Sc_{0.27}N$.

The depolarizing field lowers the potential barrier in AlN and facilitates rapid vertical growth of the nanodomains towards the bottom electrode, which occurs on a timescale much shorter than the Landau-Khalatnikov relaxation time $\tau_{LK}$~$10^{-8}$s. This is followed by a slower lateral growth of the domain stripes in the AlN film, lasting much longer than $\tau_{LK}$. The scenario of nanodomain nucleation and growth is illustrated by the images in **Fig. 5(f)**. Images $1-10$ show the distribution of $P_z$ in the cross-section of the nanostructured films at moments in time numbered from "1" to "8" during the voltage sweep shown in **Figs. 5(a).** The time dependences of the averaged polarization $\overline{P_z}$ and vertical surface displacement $u_z$, corresponding to the voltage sweep, are shown in **Fig. 5(b)** and **Fig. 5(c)**, respectively. The ferroelectric hysteresis loops of the averaged $\overline{P_z}$ and the butterfly-like loops of $u_z$ are shown in **Figs. 5(d)** and **5(e)**, respectively.

We also analyzed the polarization switching in the inverted structures, where we varied the shape of AlN clusters in the $Al_{0.73}Sc_{0.27}N$ film. Corresponding $P_3(U)$ loops are shown in **Fig. 2(e)**. Corresponding dependence of the coercive field $E_c$ on the depolarization factor $n$ is shown by the blue curve in **Fig. 3(a)**. We revealed that the dependence $E_c(n)$ is nonmonotonic; it has a rather flat minimum near $n \approx 0.5$, in contrast to the monotonic dependence $E_c(n)$ calculated for $Al_{0.73}Sc_{0.27}N$ clusters in the film AlN (shown by the red curve in **Fig. 3(a)**). The coercive field of the $Al_{0.73}Sc_{0.27}N$ film with AlN stripes (for which $n = 0$) is a little bit more than the coercive field of a pure $Al_{0.73}Sc_{0.27}N$ film. The insignificant difference in $E_c$ originates from the lateral growth of domain walls in the highly-coercive AlN stripes, which fraction is small (less than 0.1). The coercive fields of the $Al_{0.73}Sc_{0.27}N$ film with AlN clusters, whose semi-ellipsoidal cross-section has a depolarization factor $0.01 < n < 0.8$, appear smaller than the coercive field of a pure $Al_{0.73}Sc_{0.27}N$ film. At $n > 0.8$ the coercive field exceeds the value of pure $Al_{0.73}Sc_{0.27}N$ and increases monotonically with increase in $n$. The largest coercive field corresponds to the AlN/$Al_{0.73}Sc_{0.27}N$ bilayer structure (for which $n = 1$), because the switching becomes single-domain in the limiting case $n \rightarrow 1$.

The evident difference in the coercive field behavior in the inverted structure "AlN clusters – $Al_{0.73}Sc_{0.27}N$ film"(compared to the structure "$Al_{0.73}Sc_{0.27}N$ clusters – AlN film") can be explained by the significant difference in the domain nucleation conditions (compare **Fig. 3(d)** with **Figs. 3(b)-(c)**), as well as by the increasing contribution of the domain walls lateral growth at $n \geq 0.5$. The nucleation conditions in the inverted structure depend strongly on the distribution of internal electric field, shown in **Figs. 2(i)-2(k)** and **Fig. 4(d)**. The dependence of $E_c(n)$, shown by the blue curve in **Fig. 3(a)**, can be explained by the distribution of the internal electric field. In particular, AlN clusters with $0.1 < n < 0.8$ reduce the coercive field most effectively because the internal electric field becomes negative near the contact between the curved cluster surface and the top electrode, while remaining relatively large inside the cluster (see dark-blue and light-blue regions in **Fig. 2(k)**, and curve 3 in **Fig. 4(d)**). As a result, nanodomains nucleate in the less coercive $Al_{0.73}Sc_{0.27}N$ outside the AlN clusters in the immediate vicinity of cluster-electrode contacts (namely, at point 3 in **Fig. 4(b)**), where the internal field is the strongest (see **Fig. 3(d)**). The stray field is positive outside the apex of the AlN cluster and therefore cannot promote nucleation in the bulk of $Al_{0.73}Sc_{0.27}N$ (see curves 2 and 4 in **Fig. 4(d)**). In other words, AlN clusters with curved surfaces act as polar interfacial defects. For elongated AlN spikes with $n \ll 0.1$, the stray field is small in the nucleation region near the spike apex. Thus, for AlN clusters with $n \ll 0.1$, the stray field does not assist domain nucleation in $Al_{0.73}Sc_{0.27}N$. For flat AlN clusters with $n > 0.8$, the distance that domain walls must travel during lateral expansion in the highly-coercive AlN cluster becomes large. Thus, the AlN clusters with $n > 0.8$ increase the coercive field compared to $Al_{0.73}Sc_{0.27}N$. The "optimal" shape of the AlN cluster cross-section is close to a semi-circular cluster ($n \approx 0.5$). In this case, the negative depolarization field is sufficiently large at point 3 and inside the semi-circular cluster, while the distance that domain walls must travel within the highly-coercive AlN remains relatively small. Thus, clusters with a semi-circular cross-section produce the largest reduction of the coercive field (up to a factor of ~1.22), compared to other cluster shapes.

The internal field, which is polarizing inside the ferroelectric film due to the smaller spontaneous polarization of $Al_{0.73}Sc_{0.27}N$ and depolarizing in the non-switchable cluster due to the larger spontaneous polarization of AlN, lowers the potential barrier in the film and facilitates nanodomain nucleation near the polar interfacial defects (i.e., near point 3 in **Fig. 3(b)**). The next stages involve the vertical growth of needle-like nanodomains toward the bottom electrode, followed by their lateral growth in both the film and the AlN clusters due to the proximity effect. Nanodomain nucleation and subsequent vertical and lateral growth is illustrated in **Fig. 3(d)** and **6(f)**. Images $1 - 10$ in **Fig. 6(f)** show the distribution of $P_z$ in the cross-section of the nanostructures films at the corresponding stages of the voltage sweep shown in **Figs. 6(a).** The time dependences of the averaged polarization $\overline{P_z}$ and vertical surface displacement $u_z$, corresponding to the voltage sweep, are shown in **Fig. 6(b)** and **Fig. 6(c)**, respectively. The ferroelectric hysteresis loops of the averaged $\overline{P_z}$ and the butterfly-like loops of the $u_z$ are shown in **Figs. 6(d)** and **6(e)**, respectively.

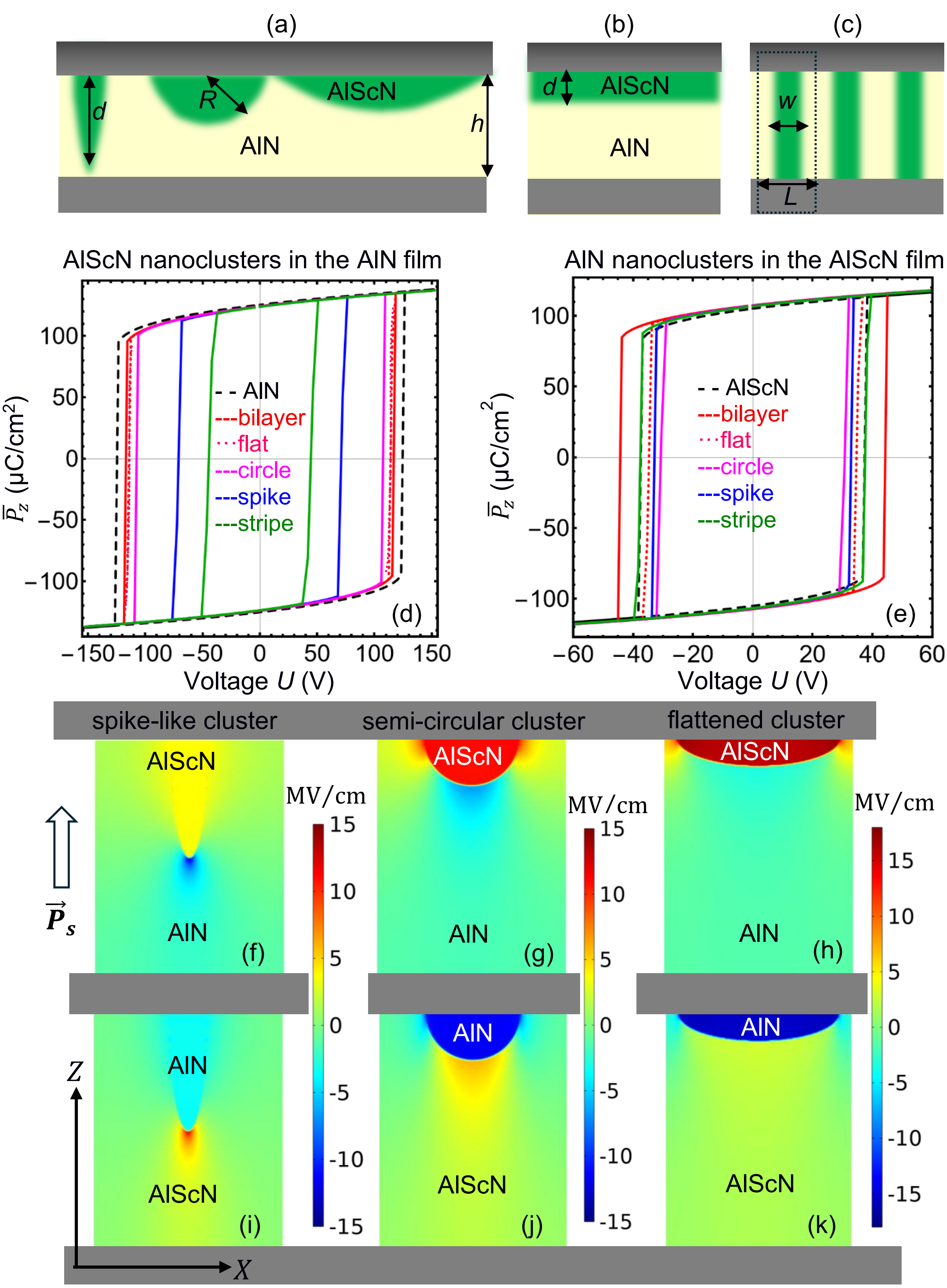

**FIGURE 2**. **(a)** Cross-sections of $Al_{1-x}Sc_xN$ clusters: spike-like, semi-circular, and flattened shape modeled by semi-ellipsoids with different aspect ratios $\frac{R}{d}$. **(b)** Horizontal $Al_{1-x}Sc_xN$/AlN bilayer and **(c)** vertical striped $Al_{1-x}Sc_xN$–AlN nanostructures. **(d)** Polarization-voltage hysteresis loops $\overline{P}_z(U)$ calculated for different cross-section shapes of $Al_{0.73}Sc_{0.27}N$ nanoclusters in an AlN film placed between parallel-plate electrodes. **(e)** $\overline{P}_z(U)$ loops calculated for different cross-section shapes of AlN nanoclusters in an $Al_{0.73}Sc_{0.27}N$ film placed between parallel-plate electrodes. Black dashed loops correspond to pure AlN (or $Al_{0.73}Sc_{0.27}N$) films, red loops correspond to 4 nm $Al_{0.73}Sc_{0.27}N$ / 46 nm AlN (or 46 nm AlN / 4 nm $Al_{0.73}Sc_{0.27}N$) bilayers, scarlet dotted loops correspond to flattened $Al_{0.73}Sc_{0.27}N$ (or AlN) clusters with sizes $R = 17.2$ nm and $d = 5.8$ nm, magenta loops correspond to semi-circular $Al_{0.73}Sc_{0.27}N$ (or AlN) clusters with cross-section radius $R = 10$ nm, blue loops correspond to spike-like $Al_{0.73}Sc_{0.27}N$ (or AlN) clusters with cross-section sizes $R = 4$ nm and $d = 25$ nm, and green loops correspond to $Al_{0.73}Sc_{0.27}N$ (or AlN) vertical stripes 1.6 nm thick in the AlN (or $Al_{0.73}Sc_{0.27}N$) film. Distribution of internal electric field in the AlN film with spike-like **(f)**, semi-circular **(g)**, and flattened **(h)** $Al_{0.73}Sc_{0.27}N$ nanoclusters calculated at $U = 0$. Distribution of internal electric field in the $Al_{0.73}Sc_{0.27}N$ film with spike-like **(i)**, semi-circular **(j)**, and flattened **(k)** AlN nanoclusters calculated at $U = 0$. The white arrow indicates the direction of spontaneous polarization. The film thickness is $h = 50$ nm, the cluster cross-section is fixed as $50\pi$ nm$^2$, the diffusion length is $\Delta = 0.5$ nm, and the lateral period of the structure is $L = 40$ nm. LGD parameters and elastic constants are listed in **Tables S1 – S2**.

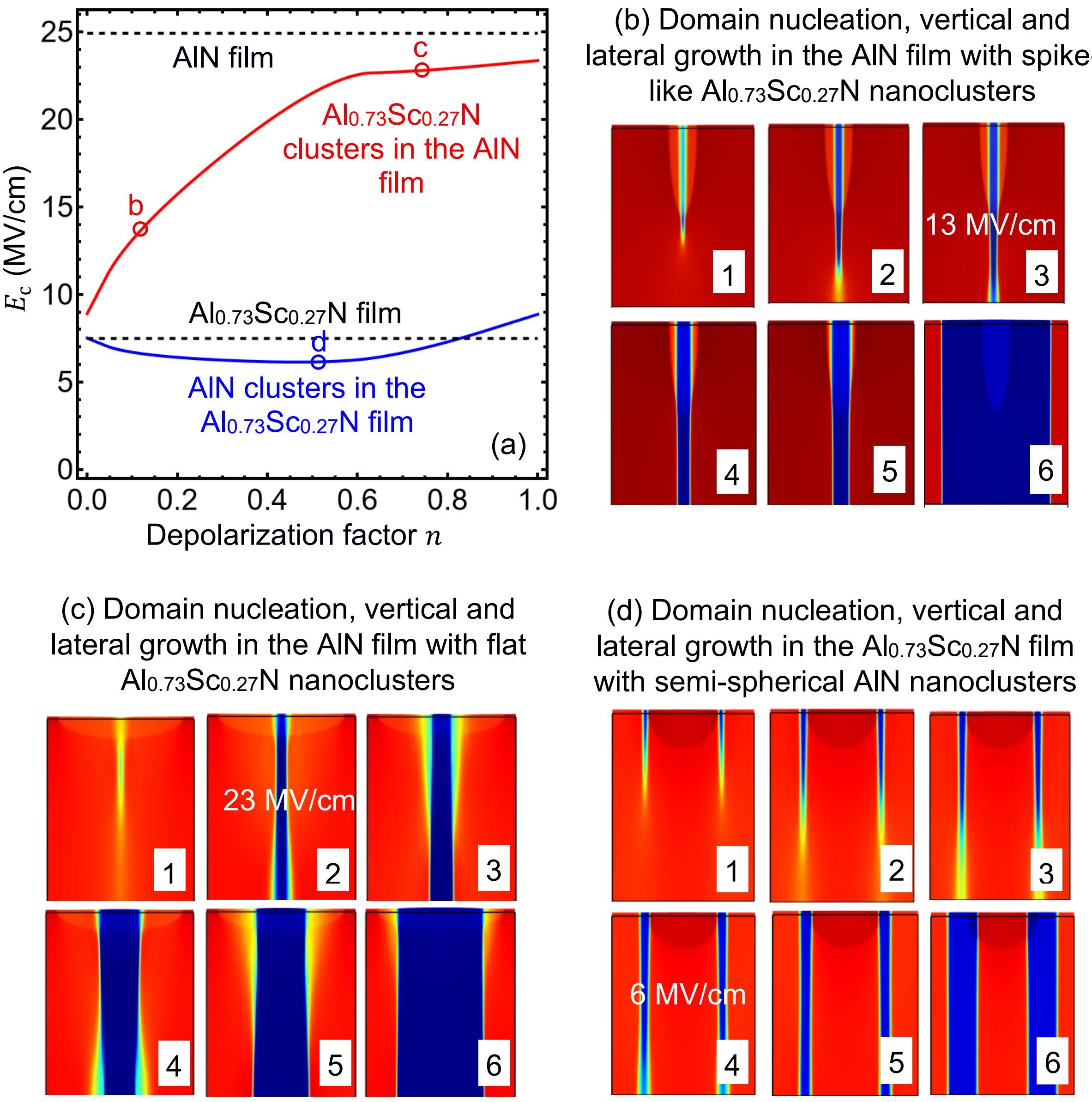

**FIGURE 3**. **(a)** Dependence of the coercive field $E_c$ on the depolarization factor $n$. The red curve corresponds to $E_c$ of the AlN film with $Al_{0.73}Sc_{0.27}N$ nanoclusters, the blue curve corresponds to $E_c$ of the $Al_{0.73}Sc_{0.27}N$ film AlN nanoclusters. Dashed horizontal lines correspond to the coercive fields of the $Al_{0.73}Sc_{0.27}N$ and AlN materials. Points "b", "c", and "d" correspond to the coercive fields in the cases shown in parts **(b)-(d)**, which are very close to the fields of the needle-like nanodomain formation (13 MV/cm, 23 MV/cm, and 6 MV/cm, respectively).

Snapshots of domain nucleation and the vertical and lateral growth stages in the AlN film with spike-like ($n = 0.14$, part **(b)**) and flattened ($n = 0.75$, part **(c)**) $Al_{0.73}Sc_{0.27}N$ nanoclusters. **(d)** Snapshots of domain nucleation, and the vertical and lateral growth stages in the $Al_{0.73}Sc_{0.27}N$ film with semi-spherical AlN nanoclusters ($n = 0.5$). The depolarization factor is $n = \frac{R}{R+d}$. The film thickness is $h = 50$ nm, the cluster cross-section is fixed as $50\pi$ nm$^2$, the diffusion length is $\Delta = 0.5$ nm, and the lateral period of the structure is $L = 40$ nm. LGD parameters and elastic constants are listed in **Tables S1 – S2**.

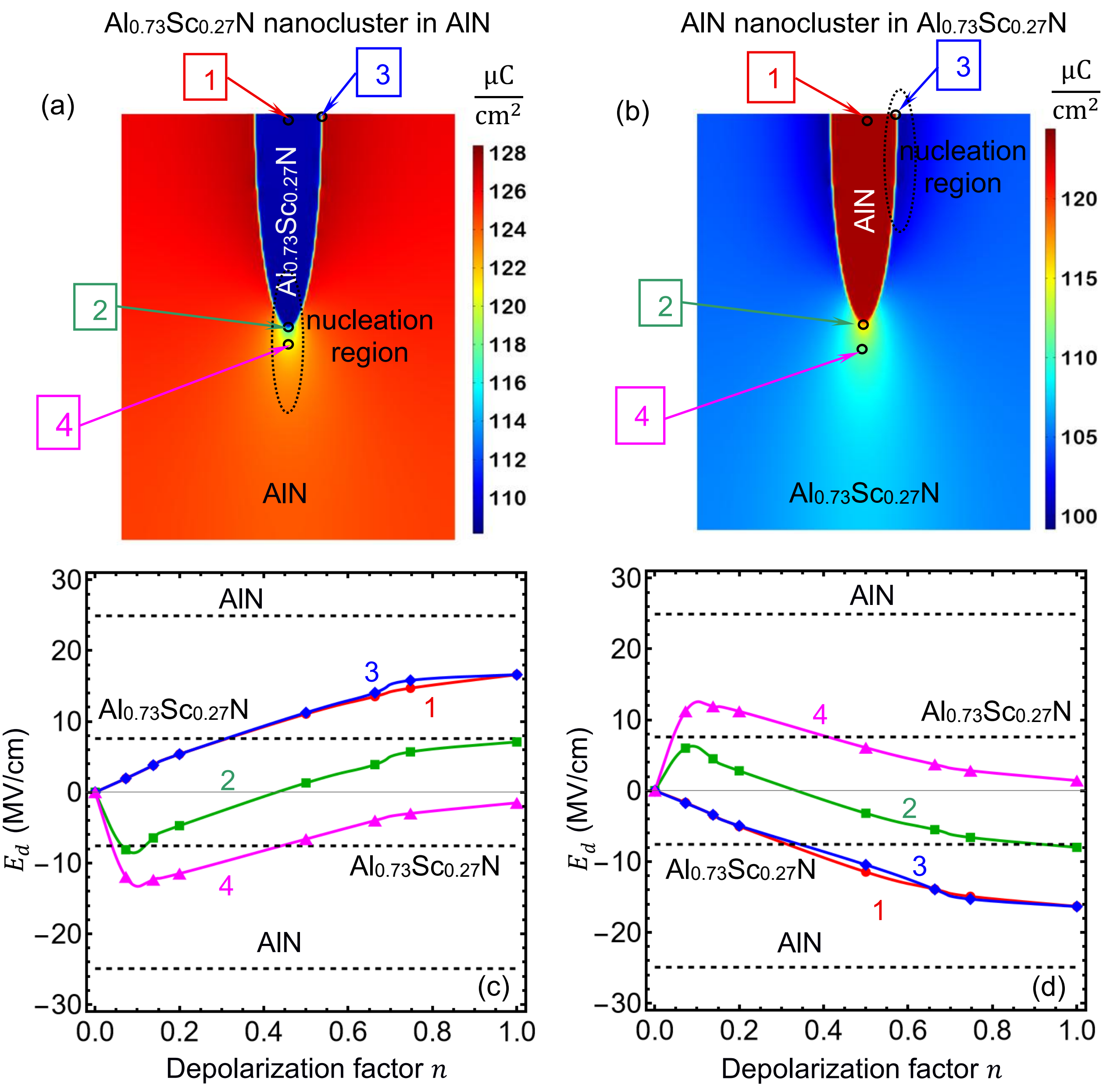

**FIGURE 4**. **(a, b)** Polarization distribution near an elongated semi-ellipsoidal cluster with selected points 1 – 4. **(c, d)** Dependence of the internal electric field $E_d$ on the depolarization factor $n$, calculated at points 1 (red curve), 2 (green curves), 3 (blue curves), and 4 (magenta curves). Dashed horizontal lines correspond to the coercive fields of the $Al_{0.73}Sc_{0.27}N$ and AlN materials. Panels **(a, c)** are calculated for $Al_{0.73}Sc_{0.27}N$ nanoclusters in the AlN film. Panels **(b, d)** are calculated for AlN nanoclusters in the $Al_{0.73}Sc_{0.27}N$ film. The applied voltage is zero ($U = 0$), and the depolarization factor $n = \frac{R}{R+d}$. The film thickness is $h$ =50 nm, the cluster cross-section is fixed as $50\pi$ nm$^2$, the diffusion length is $\Delta$= 0.5 nm, and the lateral period of the structure is $L = 40$ nm. LGD parameters and elastic constants are listed in **Tables S1 – S2**.

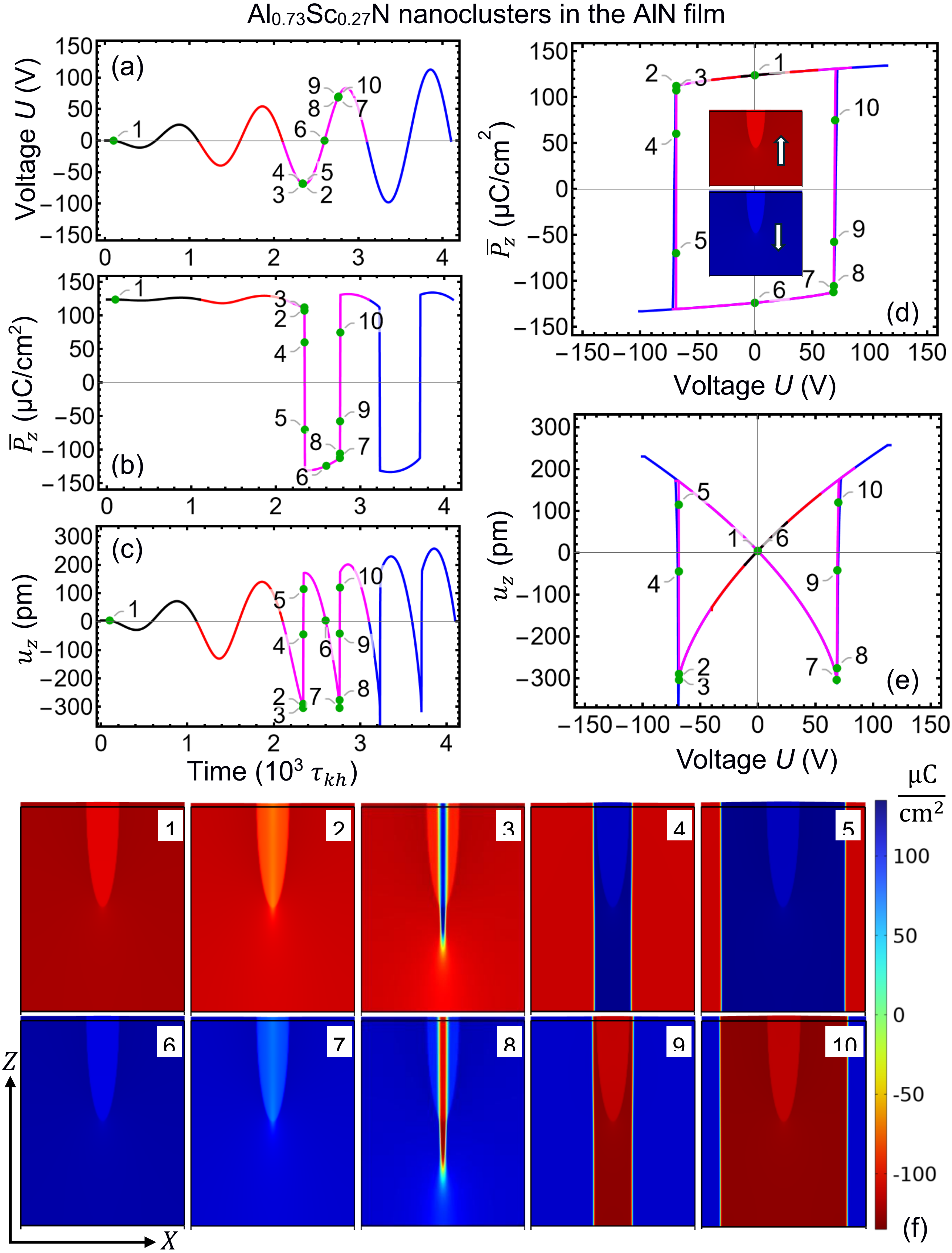

**FIGURE 5**. Time dependences of the voltage applied between the electrodes **(a)**, the average polarization $\overline{P}_z$ **(b),** and the vertical displacement $u_z$ of the top surface **(c)** for the structure consisting of spike-like $Al_{0.73}Sc_{0.27}N$ nanoclusters periodically spaced in an AlN film. Voltage dependences of the average polarization $\overline{P}_z$ **(d)** and the surface displacement $u_z$ **(e)**. **(f)** Distribution of polarization $P_z$ in the XZ-section of the structure at the time points "1" to "10" indicated in part (a). The film thickness is $h$ =50 nm, the cluster sizes are $R$ = 4 nm and $d$ = 25 nm,

the diffusion length is Δ= 0.5 nm, and the lateral period of the structure is $L$ = 40 nm. LGD parameters and elastic constants are listed in **Tables S1 – S2**.

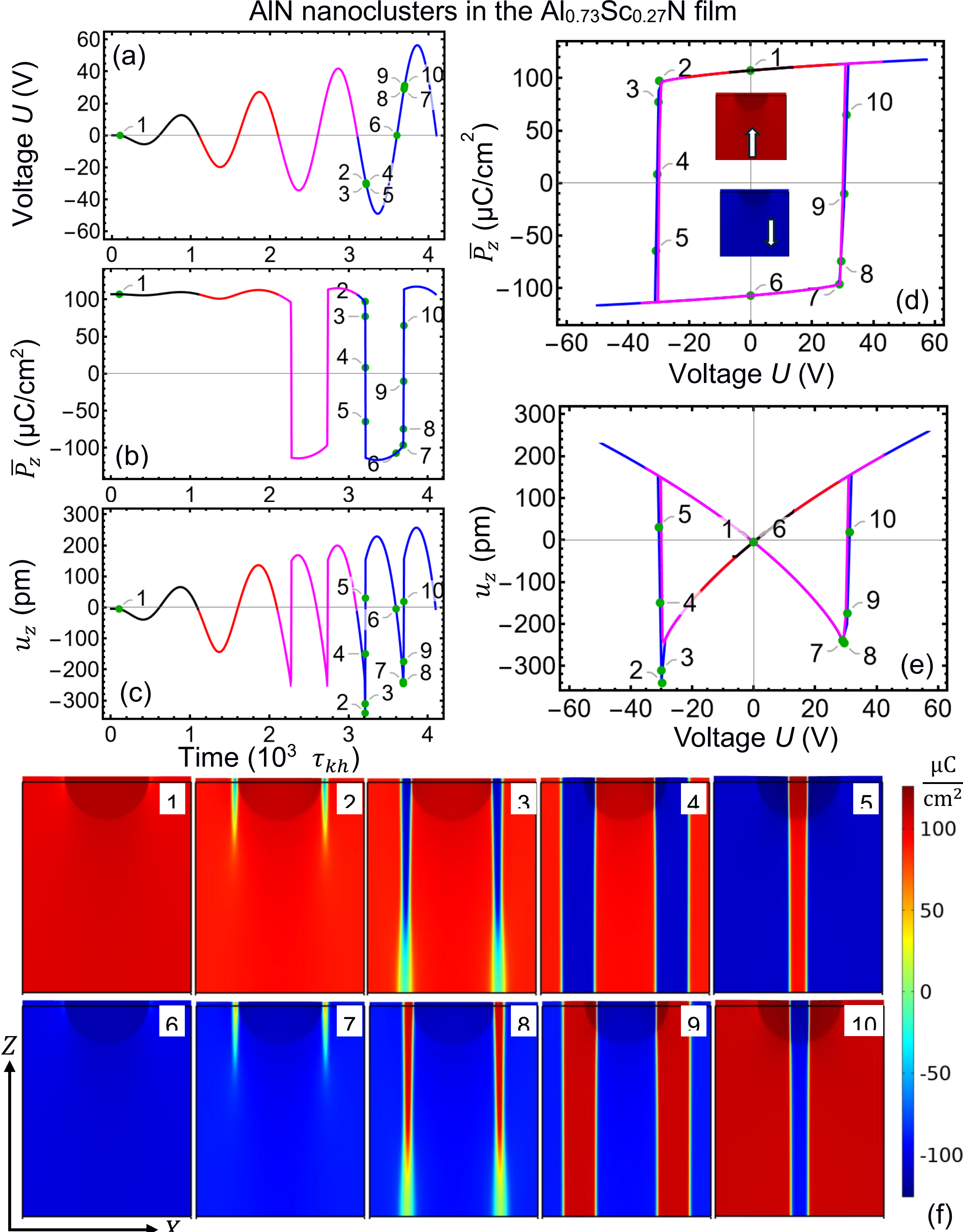

**FIGURE 6**. Time dependences of the voltage applied between the electrodes **(a)**, the average polarization $\overline{P}_z$ **(b),** and the vertical displacement $u_z$ of the top surface **(c)** for the structure consisting of AlN nanoclusters with semi-circular cross-sections periodically spaced in an $Al_{0.73}Sc_{0.27}N$ film. Voltage dependences of the average polarization $\overline{P}_z$ **(d)** and surface displacement $u_z$ **(e)**. **(f)** Distribution of polarization $P_z$ in the XZ-section of the structure at the time points “1” to “10” indicated in part (a). The film thickness is $h$ =50 nm, the cluster sizes are $R = 10$ nm and $d = 10$ nm, the diffusion length is $\Delta= 0.5$ nm, and the lateral period of the structure is $L = 40$ nm. LGD parameters and elastic constants are listed in **Tables S1 – S2**.

## 4. CONCLUSION

Using Landau-Ginzburg-Devonshire thermodynamic approach and finite element modeling, we studied the influence of nanocluster geometry on polarization switching and domain nucleation induced in otherwise non-switchable polar films due to the proximity of ferroelectric nanoclusters. The boundary of the ferroelectric nanocluster embedded in the AlN film is a compositionally graded layer, whose thickness affects the domain nucleation conditions. The volume fraction of the clusters was chosen to be the same for all cluster geometries.

We explored the underlying physical mechanisms of the proximity effects in nominally non-switchable polar films with ferroelectric nanoclusters. The internal electric field, which is depolarizing inside the AlN film (due to its larger spontaneous polarization) and polarizing in the ferroelectric $Al_{0.73}Sc_{0.27}N$ cluster (due to its smaller spontaneous polarization), lowers the potential barrier and facilitates the nucleation of nanodomains in the immediate vicinity of the cluster apex, where the uncompensated bound charge is concentrated, as well as inside the clusters. The subsequent stages, namely the instant vertical growth of nanodomains toward the bottom electrode and their lateral growth in the AlN film, occur due to the proximity effect. The calculated coercive field of the AlN film with $Al_{0.73}Sc_{0.27}N$ clusters is determined by the domain nucleation conditions at the cluster apex and by the distance between the cluster apex and the bottom electrode, which is overcome by a needle-like nucleus during its vertical growth in AlN.

We show that an appropriate choice of cluster shape allows switching of the electric polarization in the AlN film at coercive fields significantly lower than the electric breakdown field due to the proximity of ferroelectric $Al_{1-x}Sc_xN$ clusters. In particular, the inclusion of spike-like $Al_{0.73}Sc_{0.27}N$ clusters reduces the coercive field of the AlN film more than the other cluster shapes (a factor of ~2 compared to the pure AlN film). The inclusion of semi-circular $Al_{0.73}Sc_{0.27}N$ clusters leads to a smaller reduction of the coercive field, whereas flattened $Al_{0.73}Sc_{0.27}N$ clusters reduce the coercive field the least (a factor of ~1.04 compared to the pure AlN film). The cluster geometry controls both the local concentration of the depolarization field near the cluster apex and the distance that a nucleated domain must propagate through the high-coercive AlN film.

We also analyzed the polarization switching in the inverted structures, where we varied the shape of AlN clusters in the $Al_{1-x}Sc_xN$ film. In this case, the internal field is polarizing inside the $Al_{0.73}Sc_{0.27}N$ film and depolarizing in the AlN cluster. This field lowers the potential barrier in the film and facilitates the nucleation of nanodomains outside the AlN clusters, in the immediate vicinity of cluster-electrode contacts. These contacts act as polar interfacial defects. The subsequent stages are the vertical growth of needle-like nanodomains toward the bottom electrode and their lateral growth in both the film and in the AlN clusters, which occur due to the proximity effect. We show that the dependence of the coercive field on the cluster aspect ratio is nonmonotonic and has a flat maximum for a wide range of aspect ratios. Corresponding coercive fields can be smaller than the field of the pure $Al_{0.73}Sc_{0.27}N$ film, because AlN clusters with a curved surface act as interfacial polar defects. In particular, AlN clusters with a semi-circular cross-section reduce the coercive field of $Al_{0.73}Sc_{0.27}N$ the most (up to ~1.22 times) compared with all other shapes studied. The nonmonotonic dependence of the coercive field in the inverted structure is explained by the combined effect of the domain nucleation conditions at the cluster-electrode contacts and the tendency of the system to reduce the distance along the film surface, which is overcome by the domain walls during their lateral motion in the AlN clusters. Since the considered nanostructured materials can be created by implantation of Sc ions into AlN films, proximity effect could be exploited through chemically engineered nanoscale defects for realizing previously unrealizable ferroelectric technologies.

**Acknowledgements.** The work of A.N.M. is funded by the EOARD project 9IOE063 (related STCU partner project is P751c) and (in part) by the DOE Software Project on "Computational Mesoscale Science and Open Software for Quantum Materials", under Award Number DE-SC0020145 as part of the Computational Materials Sciences Program of US Department of Energy, Office of Science, Basic Energy Sciences. The work of E.A.E. is funded by the National Academy of Sciences of Ukraine and (in part) by the DOE Software Project on "Computational Mesoscale Science and Open Software for Quantum Materials", under Award Number DE-SC0020145 as part of the Computational Materials Sciences Program of US Department of Energy, Office of Science, Basic Energy Sciences. The work of L.Q.C. and V.G. are sponsored by the DOE Software Project on "Computational Mesoscale Science and Open Software for Quantum Materials", under Award Number DE-SC0020145 as part of the Computational Materials Sciences Program of US Department of Energy, Office of Science, Basic Energy Sciences. V.G. received support from the Center for 3D Ferroelectric Microelectronics Manufacturing (3DFeM2), an Energy Frontier Research Center funded by the U.S. Department of Energy, Office of Science, Office of Basic Energy Sciences Energy Frontier Research Centers program under Award Number DE-SC0021118 towards experiments on the proximity effect that motivated this study. The work of S.V.K. is supported by S.V.K. start-up funds. Results were visualized in Mathematica 14.0 [34].

**Authors' contribution.** A.N.M. and V.G. generated the research idea. A.N.M. formulated the problem, suggested mathematical model, analyzed obtained results and wrote the manuscript draft. E.A.E. wrote the codes and performed numerical modeling. S.V.K., L.Q.C., D.R.E., and V.G. worked on results explanation and manuscript improvement.

## Supplementary Materials to the Manuscript

# Embedded Ferroelectric Nanoclusters can drive Polarization Reversal in a Non-Ferroelectric Polar Film via the Proximity Effect

LGD parameters of $Al_{0.73}Sc_{0.27}N$ and AlN, used in the FEM, are listed in **Table S1.** The LGD parameters of $Al_{0.73}Sc_{0.27}N$ were determined from the experimentally measured spontaneous polarization [35, 36] and linear dielectric permittivity [37] as described in Refs. [38] and [39]. The background permittivity $\varepsilon_b$ is estimated as the square of refractive index according to Ref. [40]. Also, we assume equal gradient tensor coefficients in longitudinal and transverse directions, $g_{z,\perp}^{(i)} = 1.0 \cdot 10^{-10}$ m$^3$/F.

**Table S1.** LGD model parameters of $Al_{0.73}Sc_{0.27}N$ and AlN. Adapted from Ref. [41].

| compound | $\alpha_i$, m/F | $\beta_i$, m$^5$/(F C$^2$) | $\gamma_i$, m$^7$/(F C$^4$) | $E_c^{bulk}$, MV/cm* | $\varepsilon_b^{(i)} \approx \varepsilon_{11}^{(i)}$ |
|---|---|---|---|---|---|
| AlN layer | $-2.164 \cdot 10^9$ | $-3.155 \cdot 10^9$ | $2.788 \cdot 10^9$ | 26.1 | 4 |
| $Al_{0.73}Sc_{0.27}N$ | $-2.644 \cdot 10^8$ | $-3.155 \cdot 10^9$ | $2.788 \cdot 10^9$ | 9.3 | 3 |

* The thermodynamic coercive field of a stress-free bulk material

Electrostriction coefficients $Q_{ij}$ and elastic stiffness $c_{ij}$ tensors components of ferroelectric and nominally non-ferroelectric layers, which are regarded the same, are collected from Refs. [42, 43]. They are listed in **Table S2.**

**Table S2.** Elastic parameters of AlN and $Al_{0.73}Sc_{0.27}N$. Adapted from Ref. [41].

| parameters | $Q_{ij}$, m$^4$/C$^2$ | Ref. | $c_{ij}$, GPa | Ref. |
|---|---|---|---|---|
| AlN | $Q_{13} = -0.0087$, $Q_{33} = 0.0203$ | [42] | $c_{11} = 396$, $c_{12} = 137$, $c_{13} = 108$, $c_{33} = 373$, $c_{44} = 116$, $c_{66} = 130$ | [44] |
| $Al_{0.73}Sc_{0.27}N$ | $Q_{13} = -0.0152$, $Q_{33} = 0.0406$ | [42] | $c_{11} = 319$, $c_{12} = 151$, $c_{13} = 127$, $c_{33} = 249$, $c_{44} = 101$, $c_{66} = 84$ | [44] |

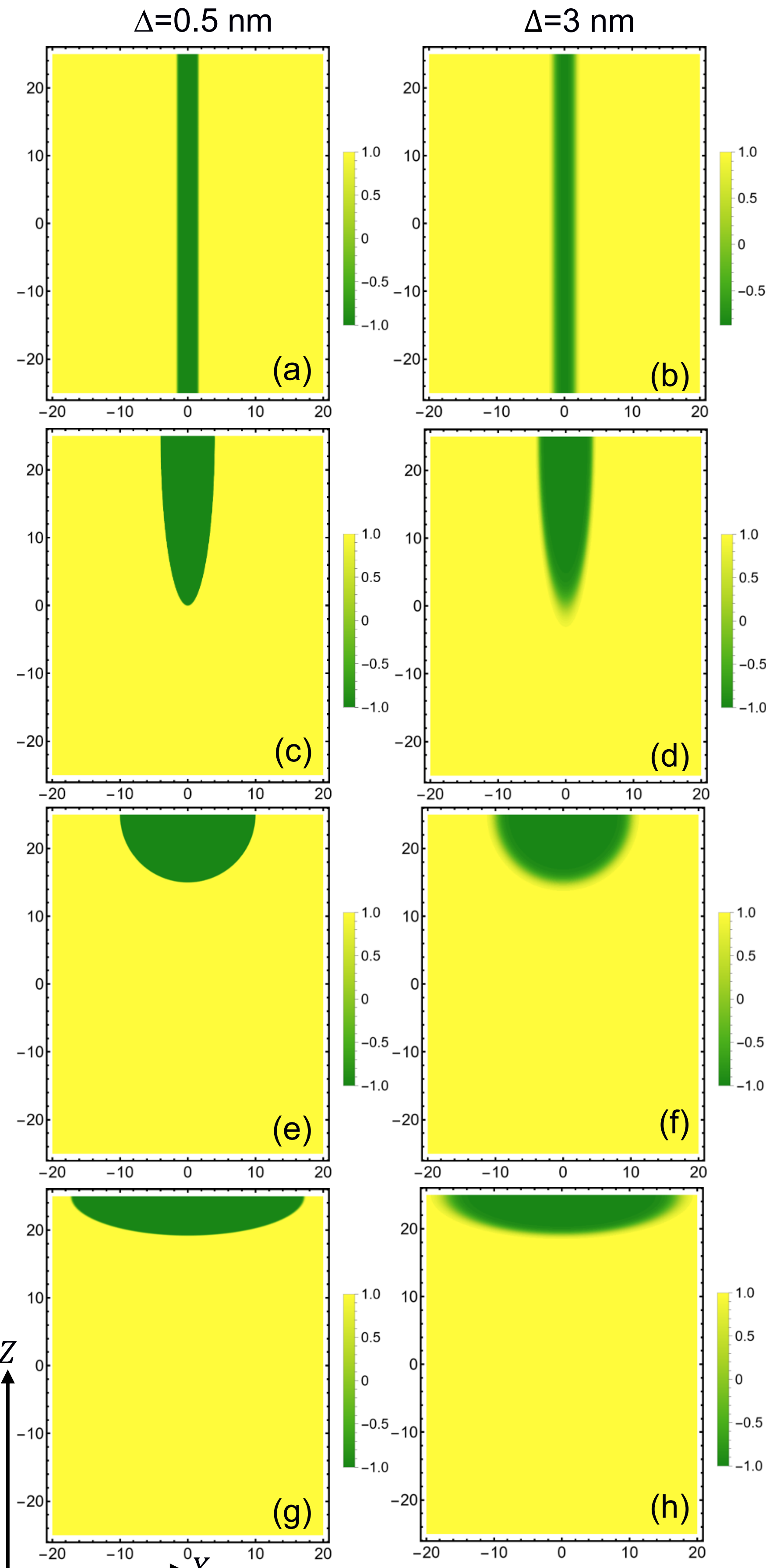

**Figure S1.** Distribution function in the XZ-section for clusters with $R = 1.6$ nm and $d \gg 50$ nm (a, b), $R = 4$ nm and $d = 25$ nm (c, d), $R = 10$ nm and $d = 10$ nm (e, f), $R = 25$ nm and $d = 4$ nm (g, h); the diffusion length $\Delta = 0.5$ (a, c, e, g) and 3 nm (b, d, f, h). The film thickness is $h$ =50 nm, and the lateral period of the structure is $L = 40$ nm.

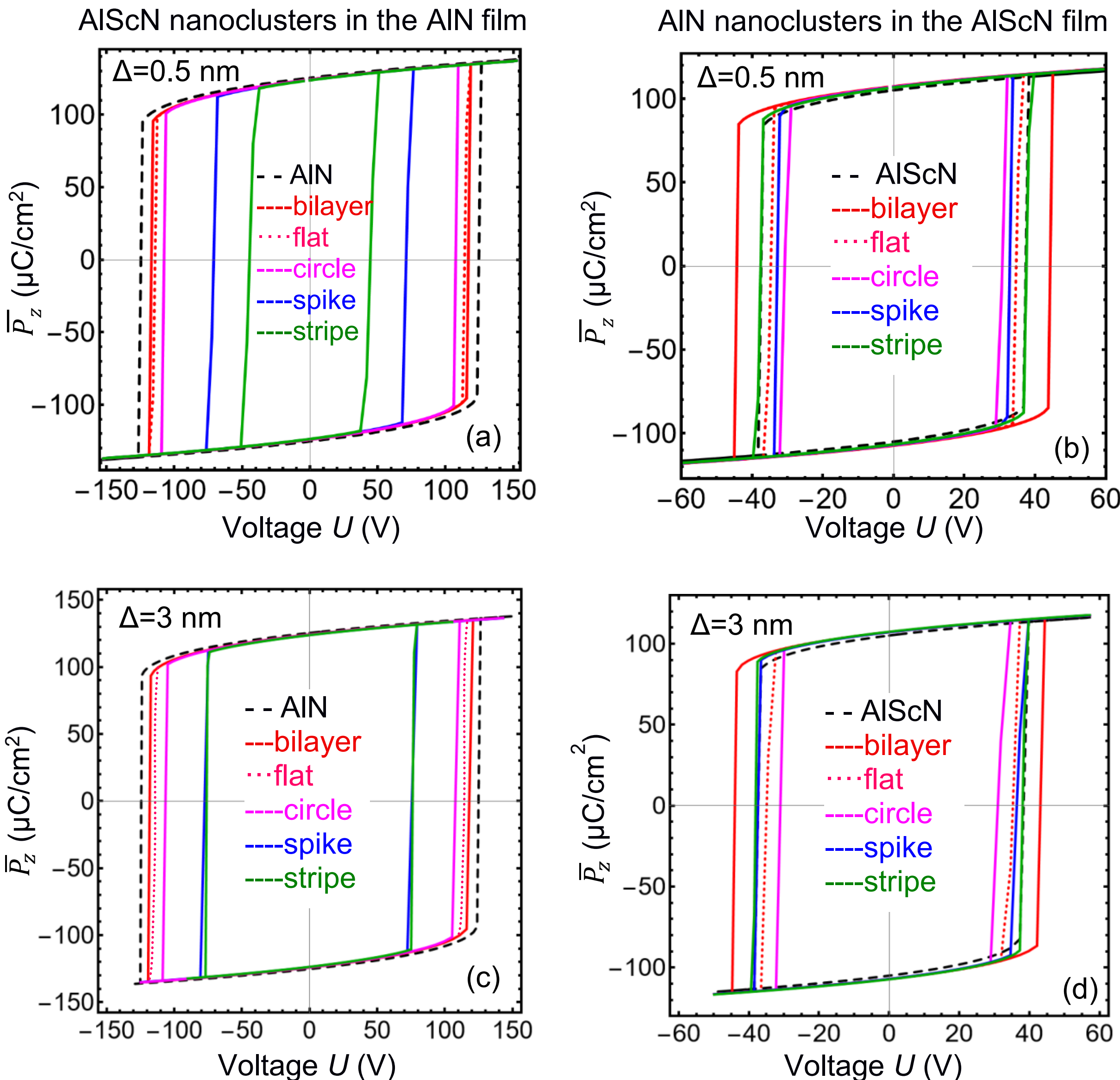

**FIGURE S2**. Polarization-voltage hysteresis loops $\overline{P}_z(U)$ calculated for different cross-section shapes of **(a, b)** $Al_{0.73}Sc_{0.27}N$ nanoclusters in the AlN film or **(c, d)** AlN nanoclusters in the $Al_{0.73}Sc_{0.27}N$ film, placed between parallel-plate electrodes. Black dashed loops correspond to the pure AlN **(a, c)** and $Al_{0.73}Sc_{0.27}N$ **(b, d)** films, red loops correspond to the 4 nm $Al_{0.73}Sc_{0.27}N$ / 46 nm AlN **(a, c)** or 46 nm AlN / 4 nm $Al_{0.73}Sc_{0.27}N$ **(b, d)** bilayers, scarlet dotted loops correspond to the flattened $Al_{0.73}Sc_{0.27}N$ **(a, c)** or AlN **(b, d)** clusters with sizes $R = 17.2$ nm and $d = 5.8$ nm, magenta loops correspond to the semi-circular $Al_{0.73}Sc_{0.27}N$ **(a, c)** or AlN **(b, d)** clusters with cross-section radius $R = 10$ nm, blue loops correspond to the spike-like $Al_{0.73}Sc_{0.27}N$ **(a, c)** or AlN **(b, d)** clusters with cross-section sizes $R = 4$ nm and $d = 25$ nm, and green loops correspond to the $Al_{0.73}Sc_{0.27}N$ **(a, c)** or AlN **(b, d)** vertical stripes of thickness 1.6 nm in the AlN **(a, c)** or $Al_{0.73}Sc_{0.27}N$ **(b, d)** film. The diffusion length is $\Delta = 0.5$ nm **(a, d)** and $\Delta = 3$ nm **(c, d)**.

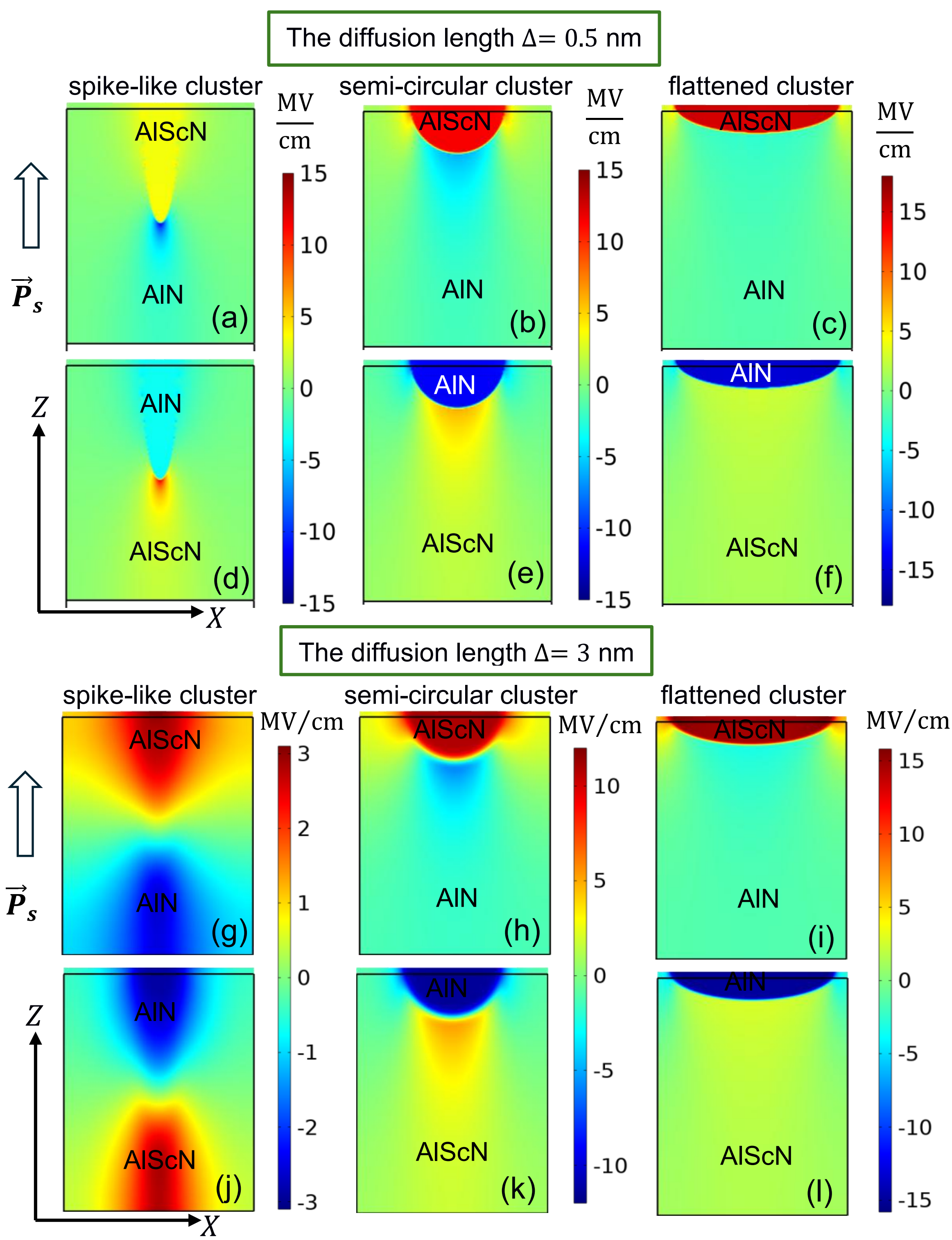

**FIGURE S3**. Distribution of the internal electric field in the AlN film with spike-like **(a, g)**, semi-circular **(b, h)** and flattened **(c, i)** $Al_{0.73}Sc_{0.27}N$ nanoclusters calculated at $U = 0$. Distribution of the internal electric field in the $Al_{0.73}Sc_{0.27}N$ film with spike-like **(d, j)**, semi-circular **(e, k)** and flattened **(f, l)** AlN nanoclusters calculated at $U = 0$. The white arrow shows the direction of spontaneous polarization. The diffusion length is $\Delta$= 0.5 nm **(a-f)** and $\Delta$= 3 nm **(g-l)**. The film thickness $h$ =50 nm, the cluster cross-section is fixed as $50\pi$ nm$^2$, and the lateral period of the structure is $L$ = 40 nm. LGD parameters and elastic constants are listed in **Tables S1** and **S2**. Black rectangles correspond to the unrelaxed state without spontaneous strain.

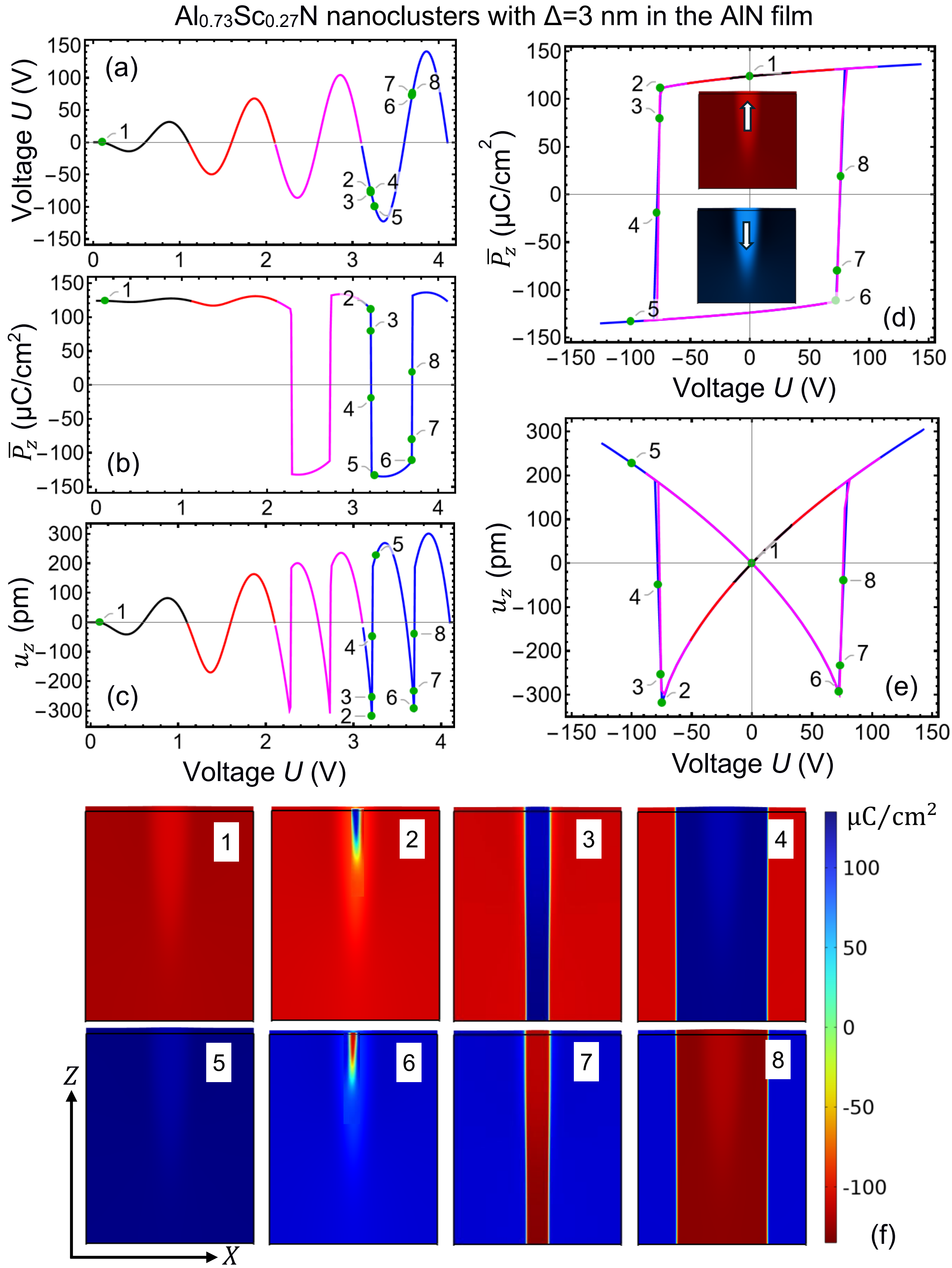

**FIGURE S4**. Time dependences of the voltage applied between the electrodes **(a)**, the average polarization $\overline{P}_z$ **(b)**, and the vertical displacement $u_z$ of the top surface **(c)** in the structure consisting of $Al_{0.73}Sc_{0.27}N$ nanoclusters with the spike-like cross-section, which are periodically spaced in the AlN film. Voltage dependences of the average polarization $\overline{P}_z$ **(d)** and surface displacement $u_z$ **(e)**. **(f)** Distribution of polarization $P_z$ in the XZ cross-section of the structure at the time points "1" to "8" indicated in part (a). The film thickness is $h$ =50 nm, the cluster sizes are $R = 4$ nm and $d = 25$ nm, the diffusion length is $\Delta = 3$ nm, and the lateral period of the structure is $L = 40$ nm. LGD parameters and elastic constants are listed in **Tables S1** and **S2**.

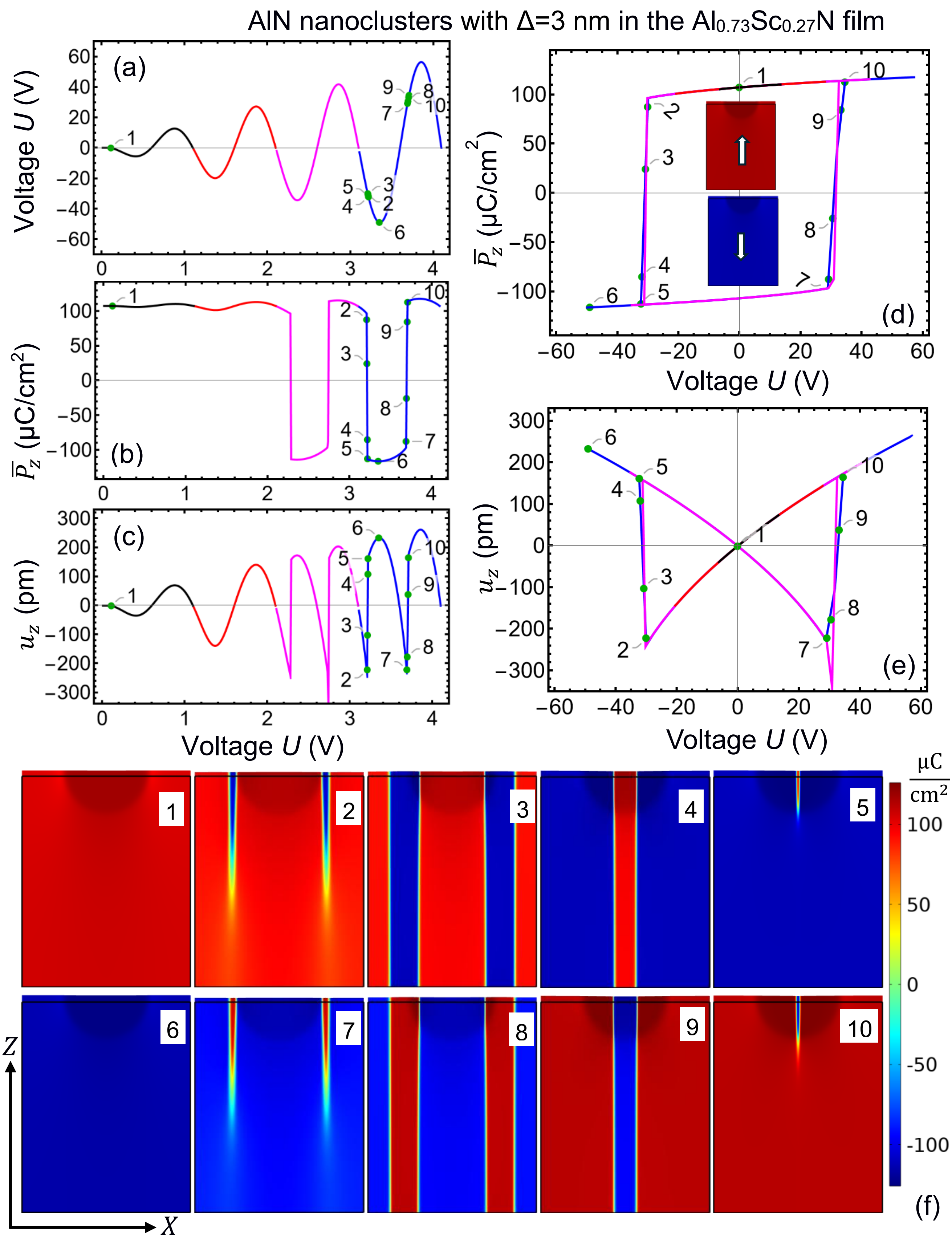

**FIGURE S5**. Time dependences of the voltage applied between the electrodes **(a)**, the average polarization $\overline{P}_z$ **(b)**, and the vertical displacement $u_z$ of the top surface **(c)** in the structure consisting of AlN nanoclusters with semi-circular cross-section, which are periodically spaced in the $Al_{0.73}Sc_{0.27}N$ film. Voltage dependences of the average polarization $\overline{P}_z$ **(d)** and surface displacement $u_z$ **(e)**. **(f)** Distribution of polarization $P_z$ in the XZ cross-section of the structure at the time points "1" to "10" indicated in part (a). The film thickness is $h = 50$ nm, the cluster sizes are $R = 10$ nm and $d = 10$ nm, the diffusion length is $\Delta = 3$ nm, and the lateral period of the structure is $L = 40$ nm. LGD parameters and elastic constants are listed in **Tables S1** and **S2**.